\documentclass[useAMS,usenatbib]{mn2e}
\usepackage{graphicx,amsmath}
\usepackage{amssymb}
\usepackage{wrapfig}
\usepackage{esint}
\usepackage{multirow}
\usepackage[british,english]{babel}

\title[Fraction of thermal Comptonized photons that impinge back]
{A Monte Carlo estimate for the fraction of thermal Comptonized photons 
that impinge back on the soft source in  neutron star LMXBs.
}

\author[Kumar and Misra]{Nagendra Kumar$^{1}$\thanks{E-mail:nagendrak@iucaa.in} and Ranjeev Misra$^{1}$\thanks{E-mail:rmisra@iucaa.in}\\
$^{1}$\textit{Inter-University Centre For Astronomy and Astrophysics, Post Bag4, Ganeshkind, 
Pune-411007, India }}

\begin{document}
%\pdfminorversion=5
%\pdfobjcompresslevel=3
%\pdfcompresslevel=9
%\DeclareGraphicsExtensions{.pdf}
%
%\onecolumn
\date{}

\pagerange{\pageref{firstpage}--\pageref{lastpage}} 

\maketitle

\label{firstpage}

\begin{abstract}
In earlier works, it was shown that the energy dependent soft time lags
observed in kHz QPOs of neutron star low mass X-ray binaries (LMXBs)
can be explained as being due to Comptonization lags provided a
significant fraction ($\eta \sim 0.2$ - $0.8$) of the Comptonized photons
impinge back into the soft photon source. Here we use a Monte Carlo scheme
to verify if such a fraction is viable or not. In particular we consider
three different Comptonizing medium geometries: (i) a spherical shell,
(ii) a boundary layer like torus and (iii) a corona on top of an accretion disk.
Two set of spectral parameters corresponding to the 'hot' and 'cold' seed
photon models were explored. The general result of the study is that 
for a wide range of sizes, the fraction lies within 
$\eta \sim 0.3$ - $0.7$, and hence compatible with the range 
required to explain the
soft time lags. Since there is a large uncertainty in the range, 
we cannot concretely rule out any of the geometries
or spectral models, but the analysis suggests that a boundary layer type
geometry with a 'cold' seed spectral model is favoured over an accretion
corona model. Better quality data will allow one to constrain the geometry
more rigorously. Our results emphasise that there is significant heating
of the soft photon source by the Comptonized photons and hence this effect
needs to be taken into account for any detailed study of these sources.
\end{abstract}

\begin{keywords}
stars: neutron -- X-rays: binaries -- X-rays: 
radiation mechanisms: thermal
\end{keywords}

\section{Introduction}

X-ray binaries are close binary systems where the compact object 
accretes  matter from a companion star via an accretion disk. The compact 
object can be either a neutron star or a black hole, while the 
companion star is a main sequence one. The   
X-ray luminosity is usually generated in the inner accretion disk near
to the compact object. 
There are two type of X-ray binaries,  high mass X-ray binaries (HMXBs) where 
the companion star is a O or B star, and low mass X-ray binaries 
(LMXBs) where the companion star is a K or M star.
The X-ray binaries are categorised into mainly two classes, transient and 
persistent,  based on their long term X-ray variability. 
They have in general two distinct spectral states, a high luminous soft state,
 which is dominated by a black body like emission, and  a 
low luminous hard state, which is dominated by  power-law emission. 
They also typically show  a third intermediate or
transitional state where the X-ray flux is highly variable
on time scale of milliseconds to seconds. The variability is sometimes of a 
 quasi-periodic nature, and are termed  as Quasi-periodic Oscillations, QPOs.
In particular, Neutron star  LMXBs have millisecond variability and 
their kHz QPOs are  positioned in definite regions on their colour-colour plots 
\citep{Altamirano-etal2008, Straaten-van der Klis-Mendez2003}.
These QPOs  occur during the soft to
hard state transition \cite[see, eg.][]{Belloni-Mendez-Homan2007}; and they
seem to have no long term correlation with X-ray luminosity \citep{Mendez-etal1999, Misra-Shanthi2004}.

Important insights into the nature of these oscillations can be obtained
by studying the fractional root mean square r.m.s. amplitude, and 
phase delay or time-lag as a function of energy which depend on the
type of QPO and typically show complex
behaviour \citep[see for review,][]{vanderKlis2006, vanderKlis2000, Remillard-McClintock2006, Tanaka-Shibazaki1996}. The energy dependence of the r.m.s and
time-lag contain clues regarding the radiative processes that are involved
in the QPO phenomena. 

Spectral fitting reveals that thermal Comptonization is the 
main radiative mechanism for hard X-ray generation in X-ray binaries. 
In this process, the seed photons are 
Comptonized by an hot thermal electron cloud or corona.
The thermal Comptonization process is generally characterised
 by three parameters, the seed photon 
source temperature $T_b$, the corona temperature $T_e$, and the 
optical depth of the medium $\tau$ or  the average number of scattering 
$<N_{sc}>$ that a photon would undergo \citep{Sunyaev-Titarchuk1980}.  
$<N_{sc}>$ depends on the geometry  of the corona and  for a given optical 
depth, it is generally difficult to compute analytically for arbitrary corona 
shapes. Since the dominant spectral component in X-ray binaries is
due to thermal Comptonization, the energy dependent r.m.s and time-lag of
the QPOs may be related to the process. Indeed,   
the energy dependence of the r.m.s and time-lag for the lower kHz QPO  can be
explained in terms of a thermal Comptonization model and moreover, such an
analysis can provide estimates of the size and geometry of the corona
\citep[][]{Lee-Miller1998,Lee-Misra-Taam2001}. In a more detailed
work \citet{Kumar-Misra2014} studied the expected 
energy dependent time lags and r.m.s for different kinds of driving
oscillations such as in the seed photon temperature or in the coronal
heating rate, while self-consistently incorporating the heating and
cooling processes of the medium and the soft photon source. They showed
that the observed soft lag for the lower kHz QPO could be obtained
only when the driving oscillation is in the heating rate of the corona
and if a substantial fraction, $\eta$ of the Comptonized photons impinge back
into the soft photon source. However, the quantitative results obtained
depends on the the specific time-averaged spectral model used for the
 analysis. Typically in the {\it Rossi X-ray Timing Experiment } (RXTE), 
Proportional Counter Array (PCA) energy band of 3-20 keV, there
are two spectral models namely the ``hot'' and ``cold'' seed photon
models which are degenerate i.e.  they both equally  fit the data of 
neutron star LMXBs
\citep{Mitsuda-etal1984, White-etal1986,Barret2001,Salvo-stella2002, Lin-Remillard-Homan2007, Cocchi-Farinelli-Paizis2011}. 
In, \citet[][hereafter Paper I]{Kumar-Misra2016}, we employed both these
spectral models to infer the size of the medium and fraction of photons 
impinging on the soft seed source, $\eta$ for different QPO frequencies of
the transient source 4U 1608-52. While both spectral models can explain the 
r.m.s  and time lag as a function of energy, the range of the size of
the medium for the hot seed photon model 0.2-2.0 kms is significantly different
than when the cold seed photon model is used, 1- 10 kms.
 Moreover, we compared the measured 
 soft lags between two broad energy bands versus kHz QPO frequency 
\citep{Barret2013}  with the model predicted ones. 
We found that the width of medium L  decreases with QPO frequency for the  
hot-seed model, but there is no such 
trend in cold-seed  one, perhaps because the allowed range of the size
is larger. For both models, we obtained the inferred ranges of L and $\eta$.
Thus, it was shown that while interpreting the time lag of the kHz QPO as
being due to Comptonization, can lead to estimates of the size of the medium,
it is necessary to have a reliable time-averaged spectral model to do so.

A generic feature of the analysis was that since the observed time lags
are soft, there needs to be a significant fraction, $\eta > 0.2$ of the
Comptonizing photons to impinge back into the soft photon source. While
in these earlier works it has been treated as a parameter, in principle,
it should be computed for a given geometry. In this work, we endeavour to
do so, by implementing a Monte Carlo method to trace the photons as they
scatter, escape from the medium and impinge into the soft photon source.
The motivation here is to compute $\eta$ as a function of size and for
different simple geometries. We will then compare the results with
the constraints obtained in Paper I, to find if any of the geometries
are more viable. We will neglect General relativistic effects and
any bulk (including orbital) motion of the Comptonizing medium.

In the next section, we briefly discuss the scheme of the Monte Carlo method 
used for the thermal Comptonization process. 
In Section 3, $\eta$ is computed for 
three different geometries of the Comptonizing system and  in Section 5, the
results are summarised and discuss.  
\section{Monte Carlo Method}

In a Monte Carlo method a photon is tracked as it enters the Comptonizing
medium and scatters multiply till it leaves the medium. The process
is repeated for a large number of photons to build up the statistics that
would give  the emergent spectrum 
as well as the direction of each outgoing photon.
The technique has been in use for several decades now \citep[for e.g.,][]{Sazonov-Sunyaev2000, Zdziarski-Pjanka2013}. 
\citet{Pozdnyakov-Sobol-Sunyaev1983} have extensively 
reviewed the Monte Carlo method for the thermal Comptonization process.
The algorithm used in this work for the Monte Carlo method in the
lab frame  has been adopted from their paper and 
the specific scheme used is from the Appendix of
\citep{Hua-Titarchuk1995}.

\begin{figure}
\centering$
\begin{tabular}{lcr}\hspace{-0.6cm}
\includegraphics[width=0.46\textwidth]{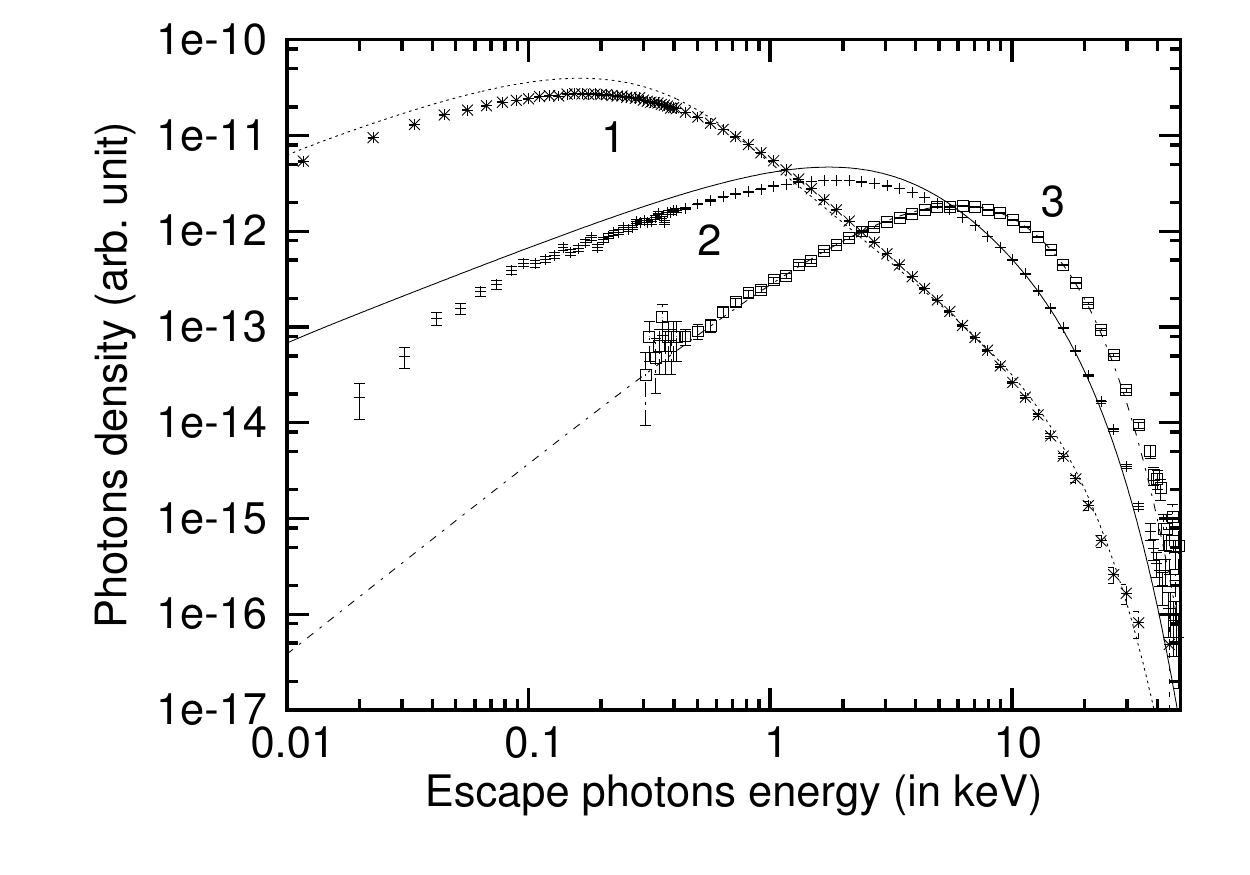}
\end{tabular}\vspace{-0.3cm}$
\caption{Spectra comparison of Monte Carlo results with analytic ones. 
Here the  points with error bars are from Monte Carlo computations
while the lines are from analytic solutions. The three curves
represent three different regimes of Comptonization namely
power-law (1), intermediate (2) and Wien peak (3). In all three regimes
the Monte Carlo technique produces results close to the analytical ones,
validating the code being used.
}
\label{MCchk}   
\end{figure}

Since our analysis is  in the non-relativistic regime, i,e. the
electron temperature, kT$_e$ and the photon energies considered are
$\ll$ m$_e$c$^2$,  and that the size of the region is much larger
than the scattering length, the diffusion limit is still valid. Thus,
we can test the code with the analytical results obtained in this
limit including the resultant spectrum from the 
Kompaneets equation \citep{Kompaneets1957}. We test the code in three
stages. First, we compute the average energy change for a monochromatic
photon of frequency $\nu$ scattering once in a thermal medium, kT$_e$
which is expected to be  $\Delta E$ = (4kT$_e$-h$\nu$)$\frac{h\nu}{m_ec^2}$.
We computed this average change in energy of the photon 
for different temperatures and found
it to match with the above expectation. Next, for a spherical geometry
we consider the average number of scatterings that a photon will undergo
 $<N_{sc}>$ and the scattering number distribution. 
For such a spherical shape geometry, one can estimate in the diffusion
limit that  $<N_{sc}>$=
$\frac{\tau^2}{2}$ and the peak of scattering distribution should be 
around $\sim$0.3$\tau^2$ 
 \citep{Sunyaev-Titarchuk1980}.  We find these expected results for the
Monte Carlo code, for e.g, for $\tau$ = 9.2,  $<N_{sc}>$ was found to be 
 41.4, and the peak of the distribution was around 25. Finally, we compare the
output spectra of the code with the analytical ones and find a good
match as shown in  Figure \ref{MCchk}. 
Here, the medium temperature is fixed at kT$_e$ 
= 3.0 keV. The  points with error-bars are from the Monte Carlo  results while
 the lines are the  analytical solutions of the 
 Kompaneets 
equation \citep[as is described in][] {Kumar-Misra2014} 
in which the ($\tau^2+\tau$) term is equated with $<N_{sc}>$.
The curve marked 1 is for the case when the soft photon
 temperature kT$_b$ = 0.1 keV and $\tau = $  9.2. For these
values the spectrum around 1 keV should be of a power-law form
and that indeed is seen. The curve marked 3 is for the case 
when the soft photon
 temperature kT$_b$ = 1.0 keV and $<N_{sc}>$ = 500. Here the emergent
spectrum is a Wien peak as expected. The curve marked 2
   is for the case when the soft photon
 temperature kT$_b$ = 1.0 keV and $\tau = $  9.2 and the
Monte Carlo spectrum matched well the the analytical one. Thus in
different regimes of Comptonization, the code gives expected 
results.

\section{Estimating the fraction $\eta_e$ for different geometries}\label{eta:}

As shown in Paper 1, an important parameter that determines
the nature of the energy dependent time lag is the fraction of photons 
impinging back
into the soft photon source. Perhaps a more physical quantity is the
fraction in terms of photon energy, i.e.
$ \eta_e =\frac{\int  n_{\gamma b} (E) E \ dE}{\int n_{\gamma} (E) E\ dE}$,
where  $n_{\gamma} (E)$ represents the photons that emerge from the
Comptonizing medium, while $n_{\gamma b} (E)$ represents those photons
which impinge back into the source. Note that this energy weigthed 
fraction $\eta_e$ would be close to the photon fraction
$ \eta =\frac{\int  n_{\gamma b} (E) \ dE}{\int n_{\gamma} (E)\ dE}$ as long
as the emergent spectrum is not highly anisotropic. In other words,
$\eta_e \sim \eta$ as long as the  spectral shape of the the photons
going into the soft source  $n_{\gamma b} (E)$ is not very different from
the average emergent spectrum $n_{\gamma} (E)$. In this section, our aim is
to  estimate $\eta_e$ for different geometries using the Monte Carlo scheme.

As mentioned earlier, the spectra of neutron star LMXBs in the 3-20 keV band
can be fitted by two degenerate models namely the ``hot'' and ``cold'' seed 
photon ones. The best fit spectral parameters for a given model, also vary 
between different observations. In Paper I, we used spectral parameters
for nine representative RXTE observations spanning a QPO frequency range
of 500 - 900 Hz. In this work we consider two sets of spectral parameters
which roughly correspond to the spectra when the QPO frequency is low ($\sim 600$ Hz and  high ($\sim 800$ Hz). This is done for both the ``hot'' and ``cold''
seed photon models. Table \ref{spec} lists the spectral parameters used
where the ``hot'' seed photon models are named Ia and Ib while the
spectra corresponding to ``cold'' seed photon model are named IIa and IIb.
The Monte Carlo computations have been done for each of these four
set of spectral parameters.

\begin{table}
%\captionsetup{width=0.3\textwidth}
%\centering
\caption{List of Comptonization spectral parameters used for the Monte Carlo
code to compute $\eta_e$. The hot and cold seed photon models are represented
by two sets of spectral parameters.
}
\label{spec} 
\begin{tabular} {lllll}
\hline
Model & index & \multicolumn{3}{l}{Comptonization parameters}\\
\cline{3-5}
& & kT$_e$ (keV) & kT$_b$ (keV) & $\tau$ \\
\hline
hot-seed & Ia & 3 & 1 & 9 \\
& Ib & 5 & 1 & 5 \\
cold-seed & IIa & 3 & 0.4 & 9 \\
& IIb & 5 & 0.4 & 5 \\
\hline
\end{tabular}
\end{table}

\subsection{Spherical/hollow shell}\label{SSHS:}

\begin{figure}%[h!]%\vspace{-1.9cm}
%\captionsetup{font=small, width=17.5cm}
\centering$
\begin{tabular}{lr}\hspace{-1.1cm} 
\includegraphics[width=0.31\textwidth]{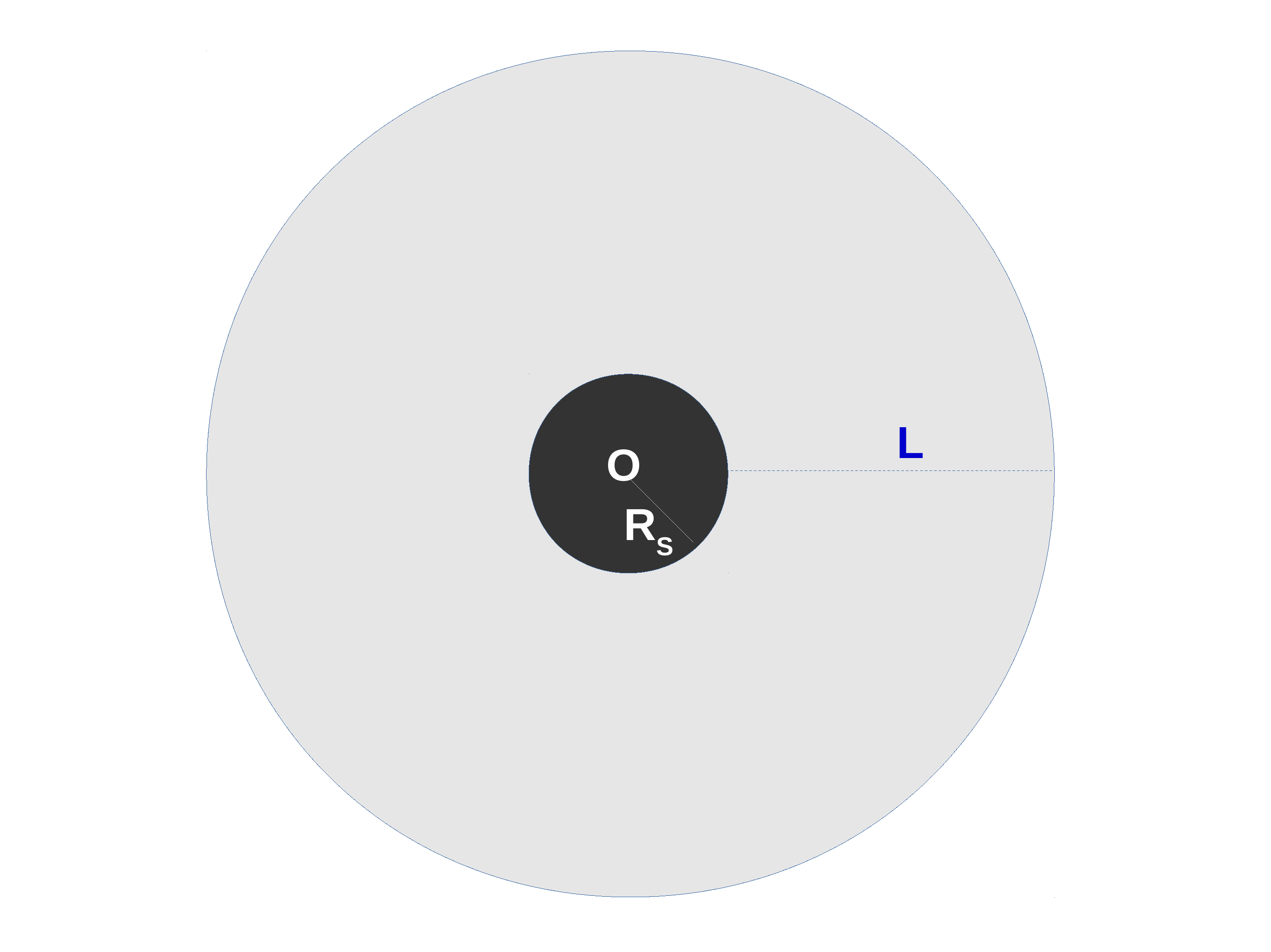} &\hspace{-1.4cm}
\includegraphics[width=0.31\textwidth]{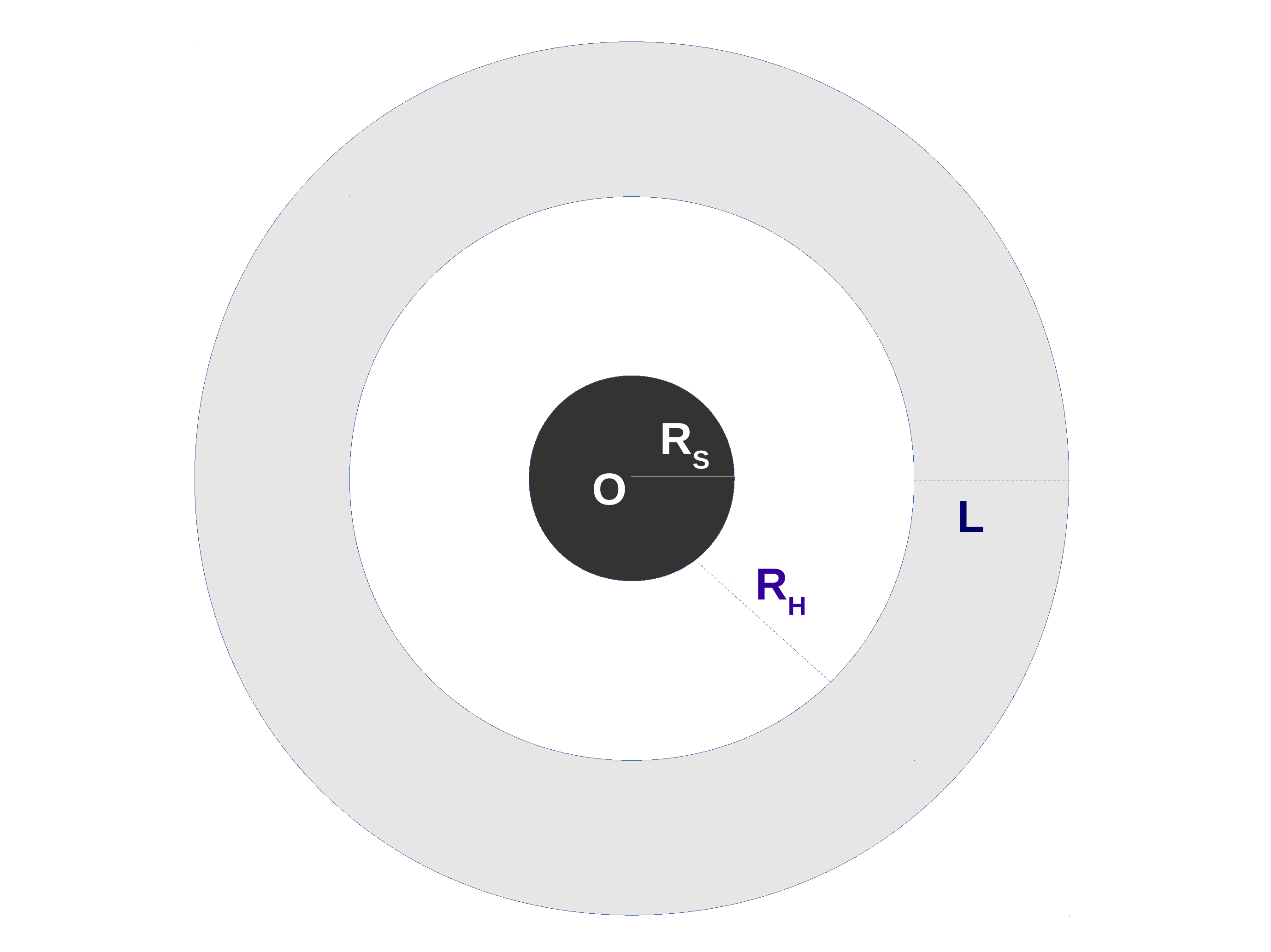} \\
\end{tabular}$%\vspace{-0.3cm}$
\caption{A cross-sectional view of the spherical shell (left panel) and 
the hollow spherical shell (right panel) geometry. The grey region 
is the Comptonizing medium of width L while the white region is the 
hollow/empty region of width R$_H$. The black region represents the
neutron star with  radius R$_S$. %{\it Nagendra remove the green marks and other labels. Keep only L, R$_H$ and R$_S$}
}
\label{SSgeo}
\end{figure}

We start with the simplest geometry depicted in Figure \ref{SSgeo},
where the neutron star is covered by a spherical shell which Comptonizes
photons from the surface of the neutron star. The radius of the neutron star
is fixed at $R_s = 10$ kms while the size of the shell $L$ is taken as
a parameter. Although perhaps not physical, we also consider for completeness,
the possibility that the Comptonizing medium is a hollow shell having a
vacant region of size $R_H$ between it and the neutron star (right
panel of Figure)

\begin{figure*}%[h!]%\vspace{-1.9cm}
%\captionsetup{font=small, width=17.5cm
\centering$
\begin{tabular}{ccc}\hspace{-0.9cm}
\includegraphics[width=0.36\textwidth]{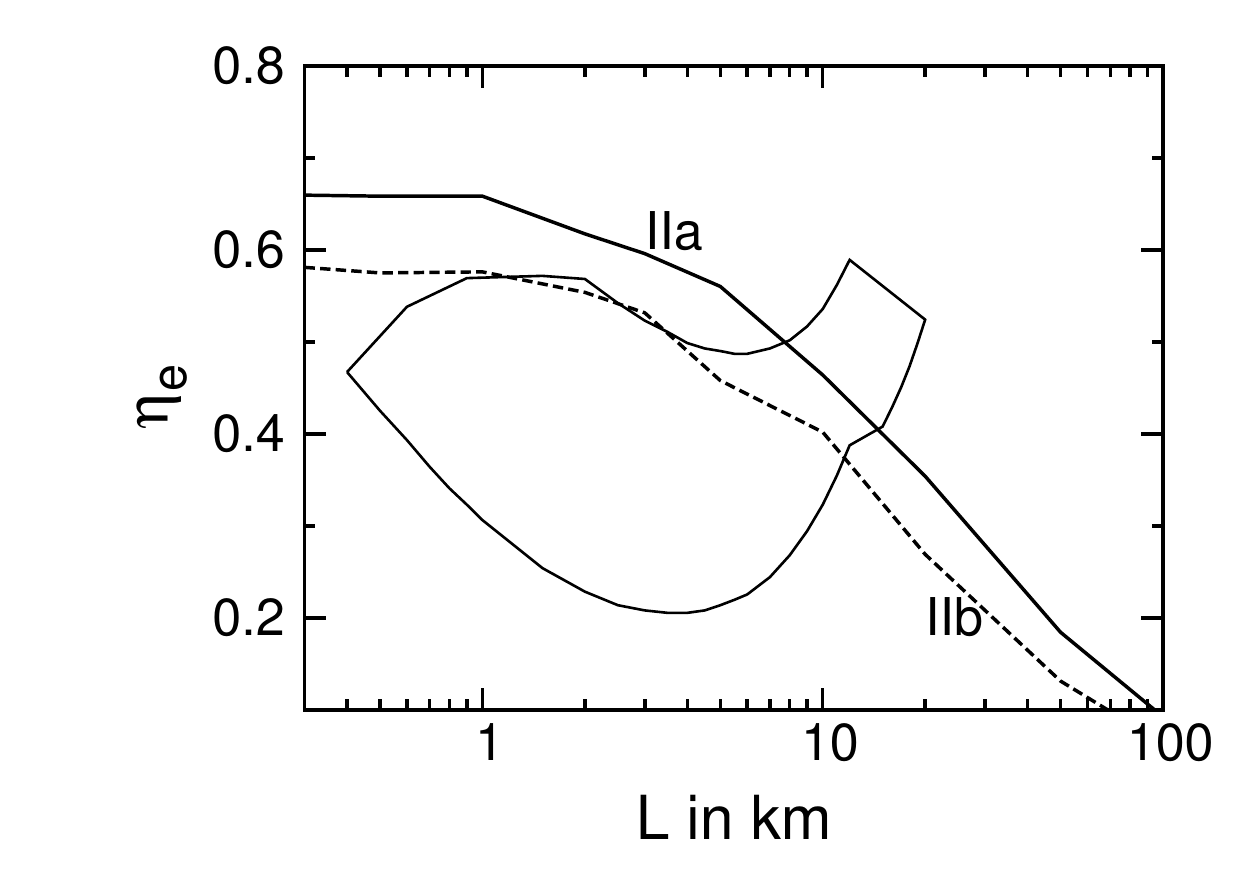} &\hspace{-1.20cm}
\includegraphics[width=0.36\textwidth]{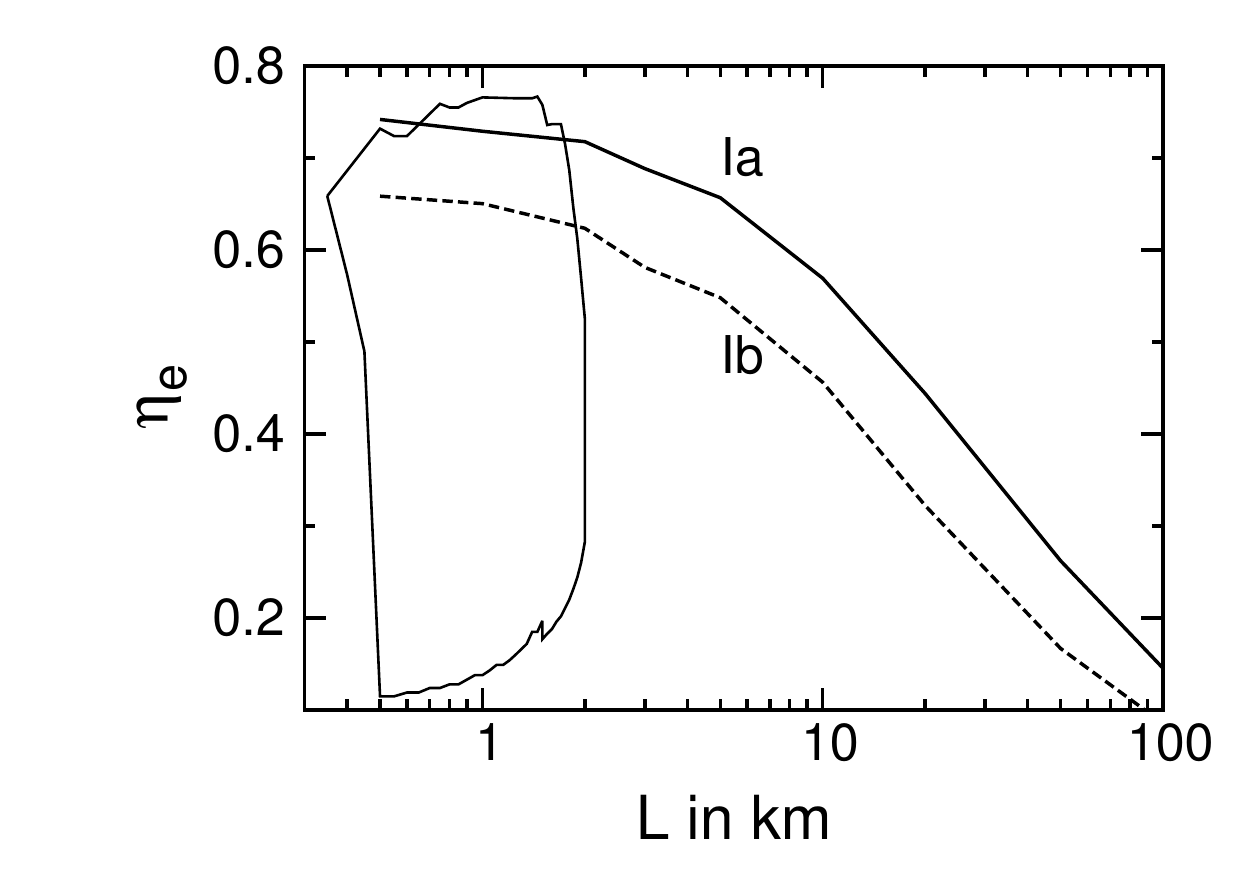} &\hspace{-1.20cm}
\includegraphics[width=0.37\textwidth]{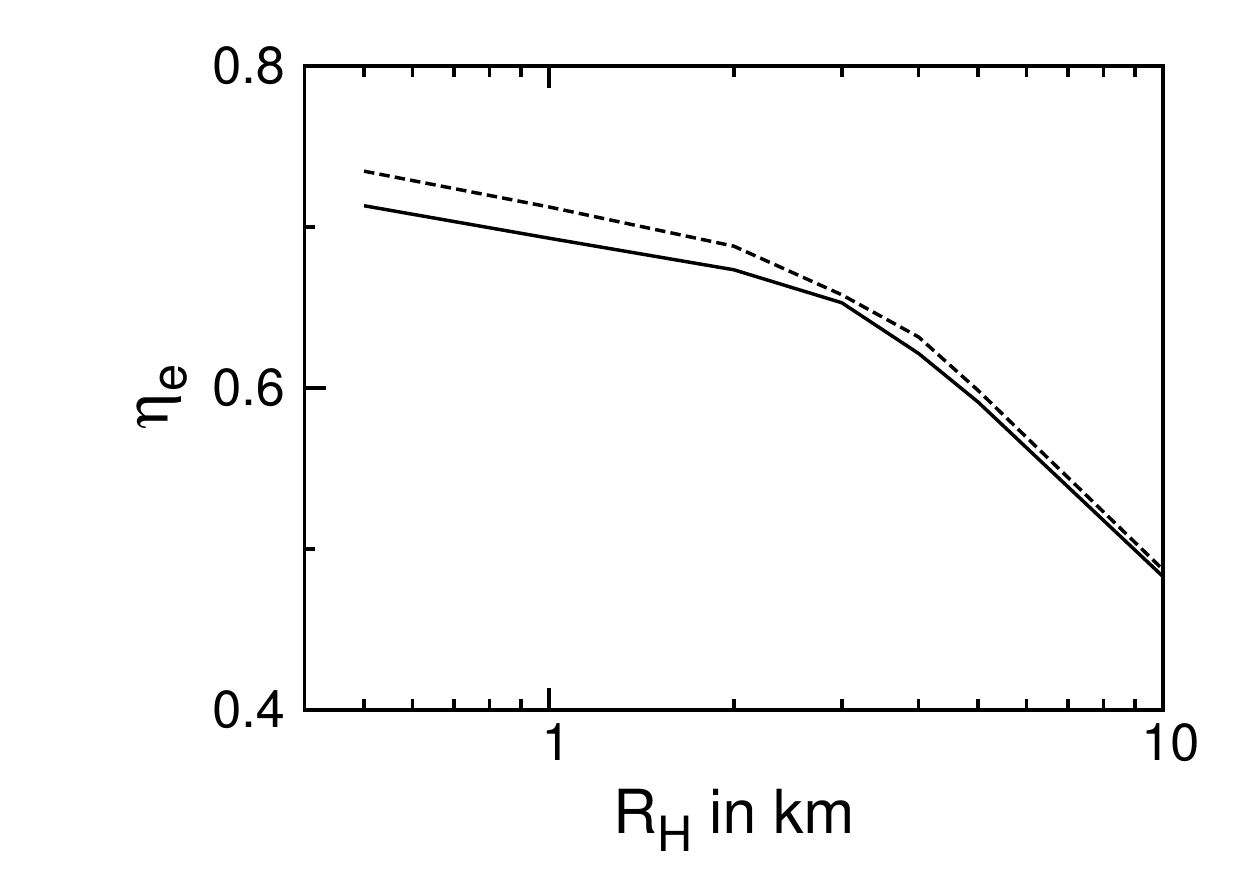} \\%\hspace{-0.90cm}
\end{tabular}$%\vspace{-0.3cm}
\caption{
Variation of $\eta_e$ versus size for the spherical shell
geometry. The left and middle
panels show $\eta_e$ variation with the size of the Comptonizing medium
for the 'cold' seed photon model (Left Panel) and for the
'hot' seed photon model (Middle Panel). The marking on the
curves (Ia,Ib,IIa,IIb) represent the spectral parameters used for
the computations as listed in Table \ref{spec}. The closed curves
represent the estimated range of $\eta$ and size that are required to explain the soft time lag in the kHz QPO \citep{Kumar-Misra2016}. 
The right panel shows the
variation of $\eta_e$ with the gap size $R_H$ for the hollow
shell model for the hot seed photon model spectra Ib. The solid
line is for $L = 1$ km while the dashed one is for  $L = 0.5$ km.
%{\it Nagendra remove the tau total and the horizontal lines from right panel}
}
\label{SSeta}
\end{figure*}

In the Monte Carlo code, a photon is released from the surface
of the neutron star and is tracked till it either escapes or impinges back
to the surface. One expects that the fraction $\eta_e$ will decrease
with increasing $L$ since the probability that a photon gets absorbed by
the surface decreases. This is indeed the case as shown in  Figure \ref{SSeta} where
the left and  middle panels show the computed $\eta_e$ as a function
of $L$ for the four spectral parameters tabulated in Table \ref{spec}.
For comparison, the plots also show the range of $\eta$ and
$L$ for the ``cold'' and ``hot'' seed photon models inferred by the
energy dependent r.m.s and time-lag of the kHz QPO (Paper I).
Although the range of $\eta$ and $L$ are rather large due to the
quality of the data, it is heartening to see that for this geometry
the computed $\eta_e$ fall within this range. If better quality data
indicate a smaller $\eta_e$ then perhaps such a geometry can be ruled
out. Naturally, a
hollow geometry would lead to lower values of $\eta_e$.
The right panel of Figure \ref{SSeta} shows this decrease of $\eta_e$ versus
the gap size $R_H$ 
for fixed values of $L = 0.5$ (solid line) and $1.0$ (dashed line) kms.

\subsection{Boundary layer  geometry} \label{BLDK:}

\begin{figure}%[h!]%\vspace{-1.9cm}
%\captionsetup{font=small, width=17.5cm}
\centering$
\begin{tabular}{lr}\hspace{-0.2cm}
\includegraphics[width=0.25\textwidth]{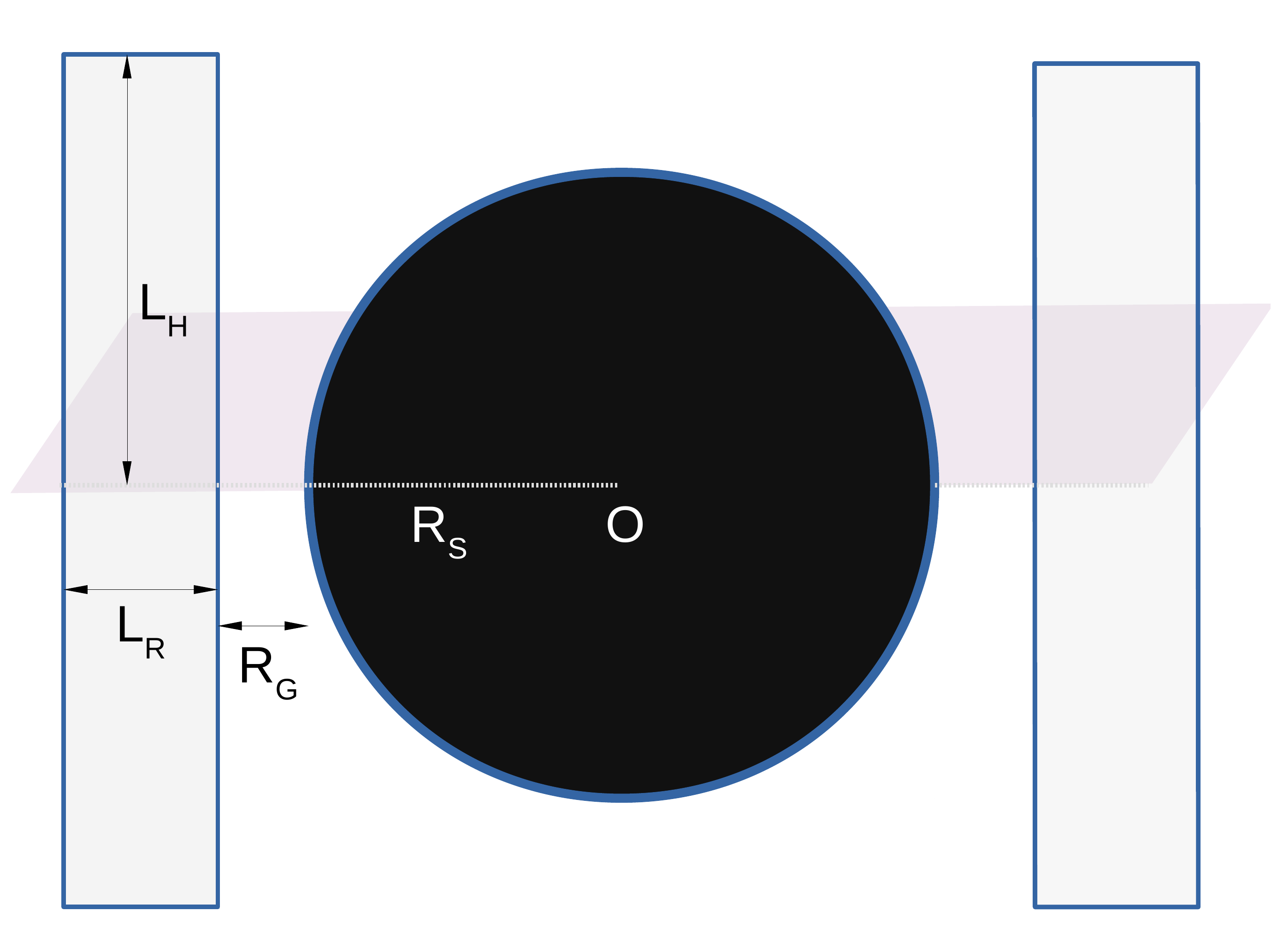} &\hspace{-0.70cm}
\includegraphics[width=0.25\textwidth]{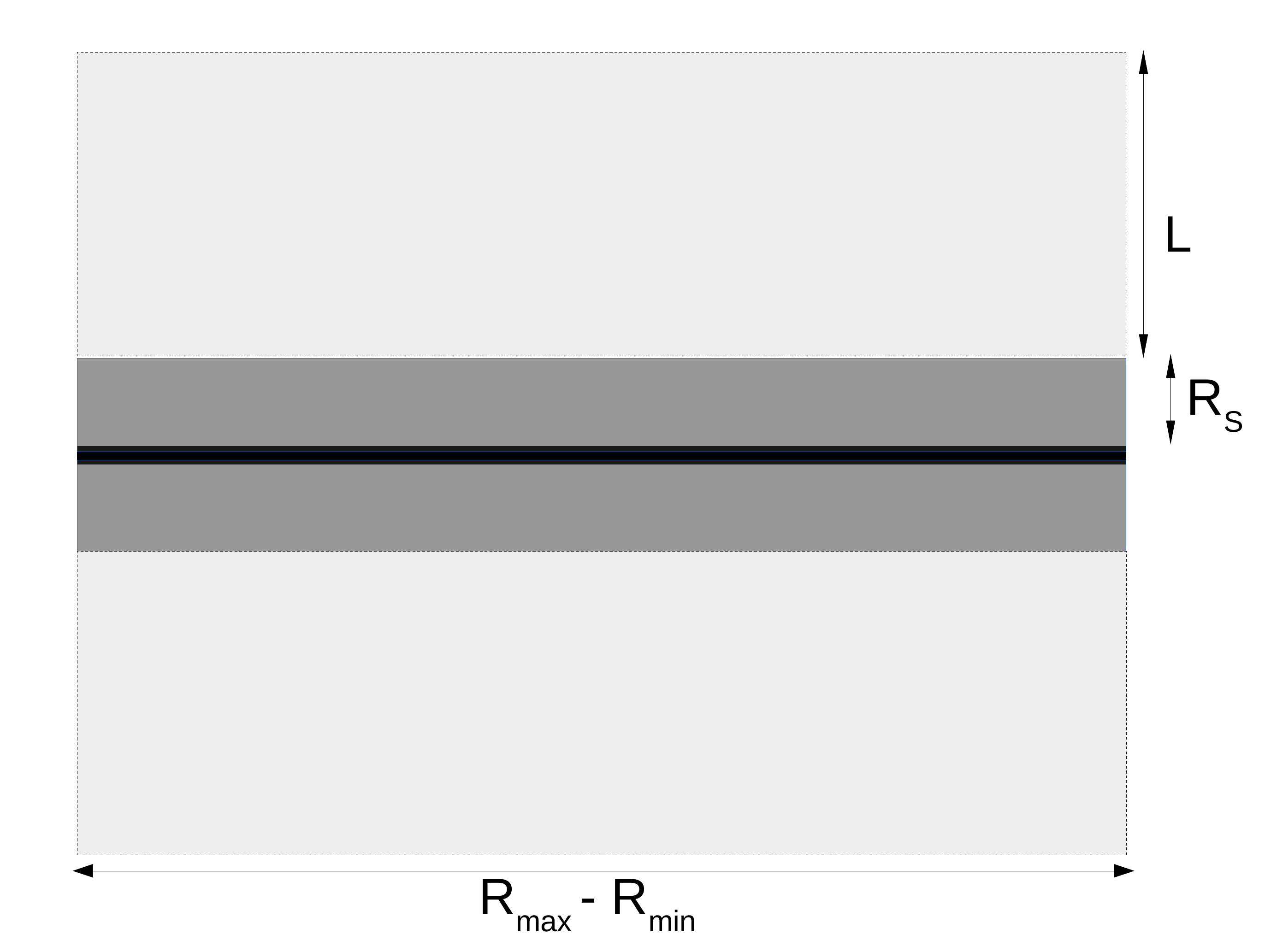} \\
\end{tabular}$%\vspace{-0.3cm}
\caption{ A cross-sectional view of the boundary layer geometry (left panel)
and the accretion disk/corona geometry (Right panel). For the boundary
layer geometry the Comptonizing medium is assumed to be a torus with
a rectangular cross-section (light grey) surrounding the neutron star
(black). For the accretion disk corona geometry the Comptonizing medium
lies above and below the accretion disk. 
%{\it Nagendra remove all markings like A B and angle lines from the left figure}.
}
\label{blgeo}
\end{figure}

The  boundary layer  is a region  that connects the accretion disk 
to the neutron star surface, i.e. the accreting material makes a 
transition from centrifugal  to pressure support near the star 
\citep[e.g.][]{Popham-Narayan1995,Popham-Sunyaev2001}. Here, we
approximate the geometry as shown in the left panel of Figure
\ref{blgeo}. We consider a rectangular torus surrounding the 
spherical neutron star. The radius of the neutron star is kept
fixed at $R_s = 10$ kms. The gap between the torus and the
neutron star $R_g$ is also fixed at a small distance of $50$ m following
\cite{Babkovskaia-Brandenburg-Poutanen2008} who estimate that the maximum
distance between the star surface and the layer is about $100$ m.
The width of the torus in the radial direction is taken to be a 
parameter $L_R$ while its half-height in the vertical direction is
$L_H$. This geometry allows for two definitions of the optical depth
$\tau$, one along the vertical and other in the radial direction and
we do the analysis for both definitions. For the same optical
depth defined in either fashion,  the average number of
scatterings $<N_{sc}>$ is smaller by a factor of $\sim 1.5$ than for
the spherical shell case studied above. Hence we use a slightly higher
values of $\tau$, 10.4 and 5.8 instead of 9 and 5 mentioned in Table
\ref{spec}. We emphasise that these changes have little effect on
the overall results.

\begin{figure}%[h!]%
%\vspace{-2.3cm}
\hspace{-0.60cm}
%\captionsetup{font=small, width=17.5cm}
%\centering
\includegraphics[width=0.5\textwidth]{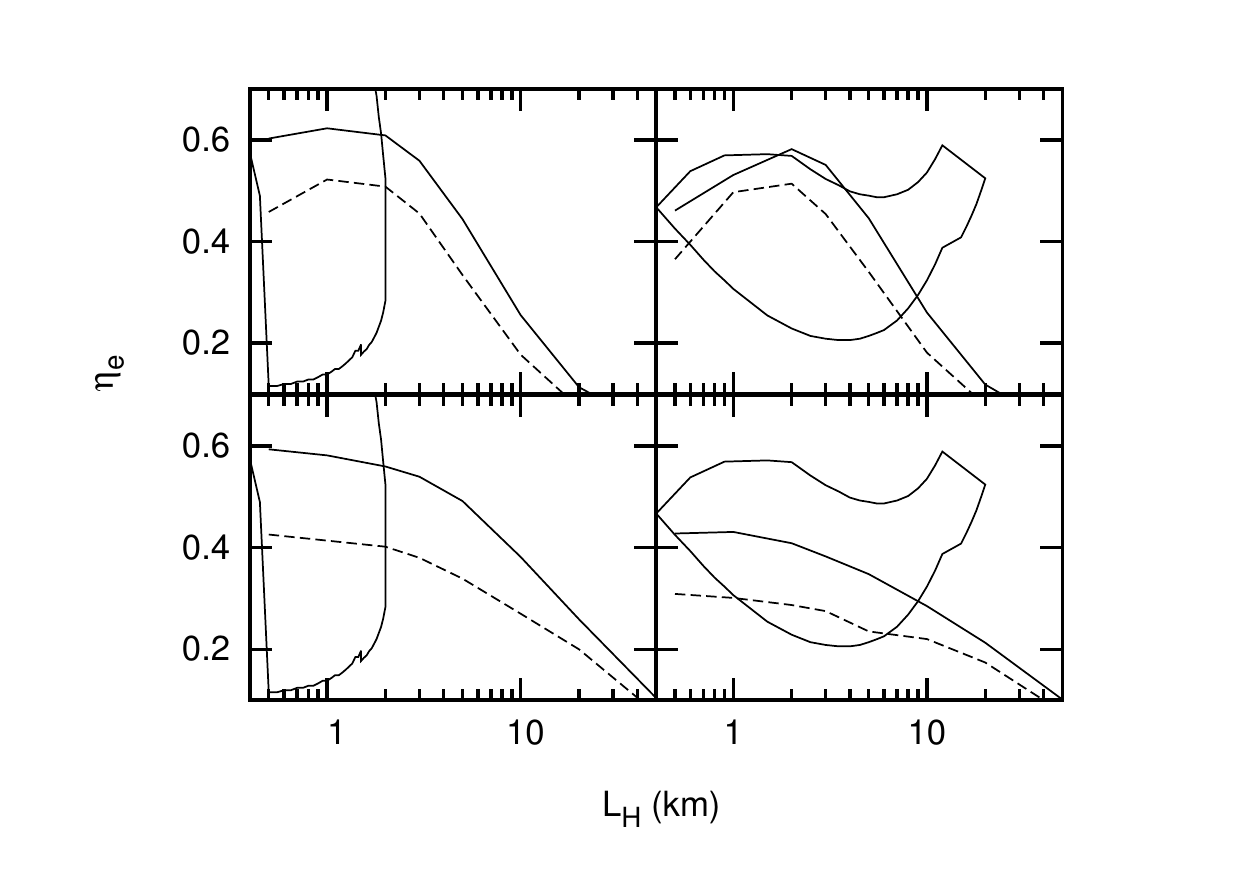} %\\
\vspace{-0.20cm}
\caption{$\eta_e$ as a variation of size for the Boundary layer geometry
when the optical depth is defined along the vertical direction. The top
and bottom two panels are for the case when the horizontal width is
taken to be  $L_R =$ 1 and 20 kms respectively. The left and right panels
are for the hot and cold seed photon models. The solid and dashed lines
are for two corresponding spectral parameters. The closed curves
show the allowed range of $\eta$ and size obtained by \citep{Kumar-Misra2016}.
%{\it Nagendra make the changes for this figure.}
}
\label{blvtt}
\end{figure}

\begin{figure}%[h!]%
%\vspace{-2.3cm}
\hspace{-0.60cm}
%\captionsetup{font=small, width=17.5cm}
%\centering
%\begin{tabular}{lr}\hspace{-0.98cm}
%\includegraphics[width=0.325\textwidth]{blrad_avs.pdf} &\hspace{-1.80cm}
\includegraphics[width=0.5\textwidth]{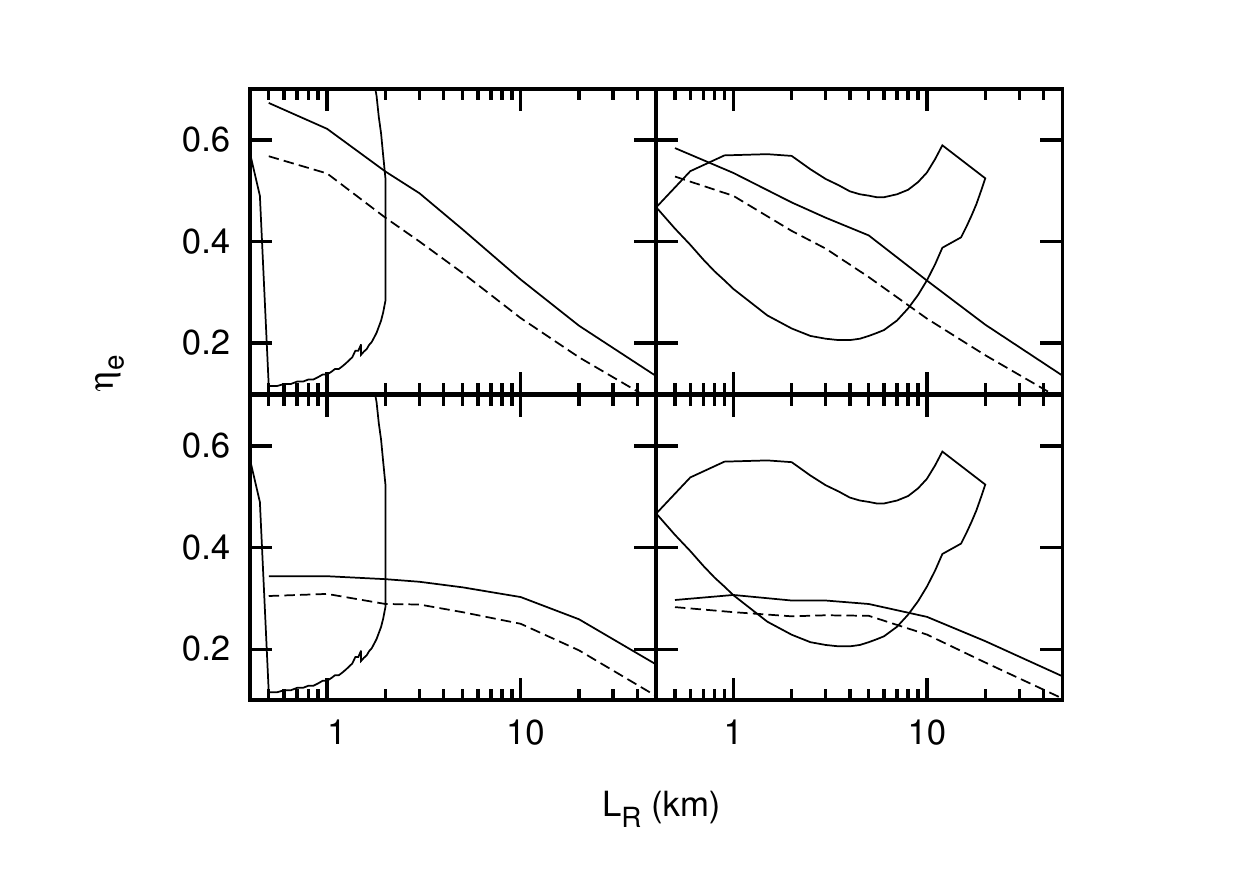} %\\
%\end{tabular}$%
\vspace{-0.20cm}
\caption{$\eta_e$ as a variation of size for the Boundary layer geometry
when the optical depth is defined along the radial direction. The top
and bottom two panels are for the case when the vertical height is
taken to be  $L_H =$ 1 and 20 kms respectively. The left and right panels
are for the hot and cold seed photon models. The solid and dashed lines
are for two corresponding spectral parameters. The closed curves
show the allowed range of $\eta$ and size obtained by \citep{Kumar-Misra2016}.
%{\it Nagendra make the changes for this figure.}    
}
\label{blrad}
\end{figure}

We first consider the case when the optical depth is defined along
the vertical direction and we explore the variation of $\eta_e$ with the
vertical height $L_H$ for fixed radial extent $L_R$ and for different spectral
parameters. This is shown in Figure \ref{blvtt} where the top panels are
for $L_R = 1$ km while the bottom ones are for  $L_R = 20$ kms. The left panels
are for the 'hot' seed photon model while the right ones are for 
the 'cold' seed photon one. The contours mark the estimated ranges of
$\eta$ and $L$ from Paper I. Figure \ref{blrad} is same as Figure \ref{blvtt}
except that now the optical depth is defined along the horizontal direction
and the top and bottom panels are for fixed values of  $L_H = 1$ and $20$ kms
respectively. It is clear from both these Figures that for a wide range
of sizes and spectral models, the fraction $\eta_e$ falls within the range
required to explain the energy dependent r.m.s and time-lags of the kHz QPOs.

\begin{figure}%[h!]%\vspace{-1.9cm}
%\captionsetup{font=small, width=17.5cm}
\centering$
\begin{tabular}{cc}\hspace{-0.90cm}
\includegraphics[width=0.28\textwidth]{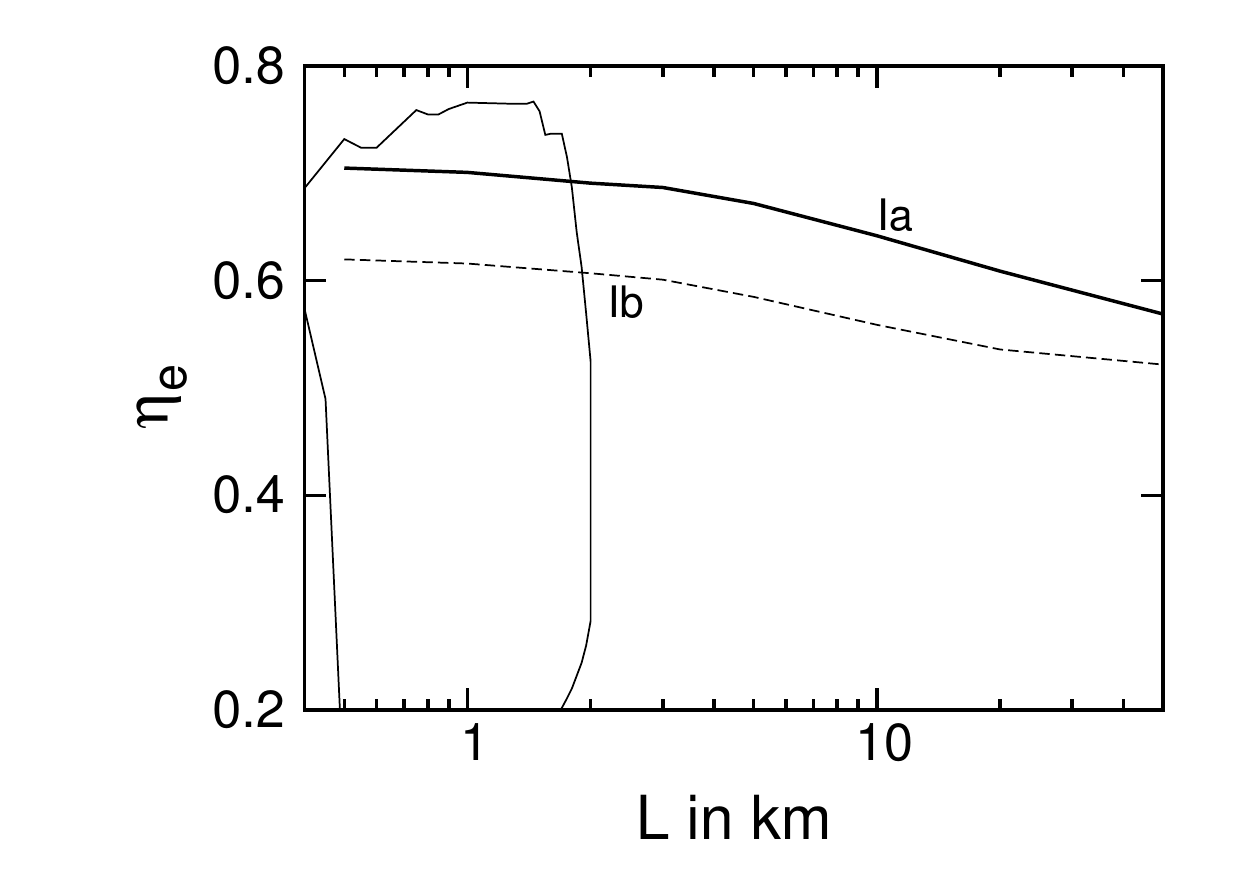} &\hspace{-1.20cm}
\includegraphics[width=0.28\textwidth]{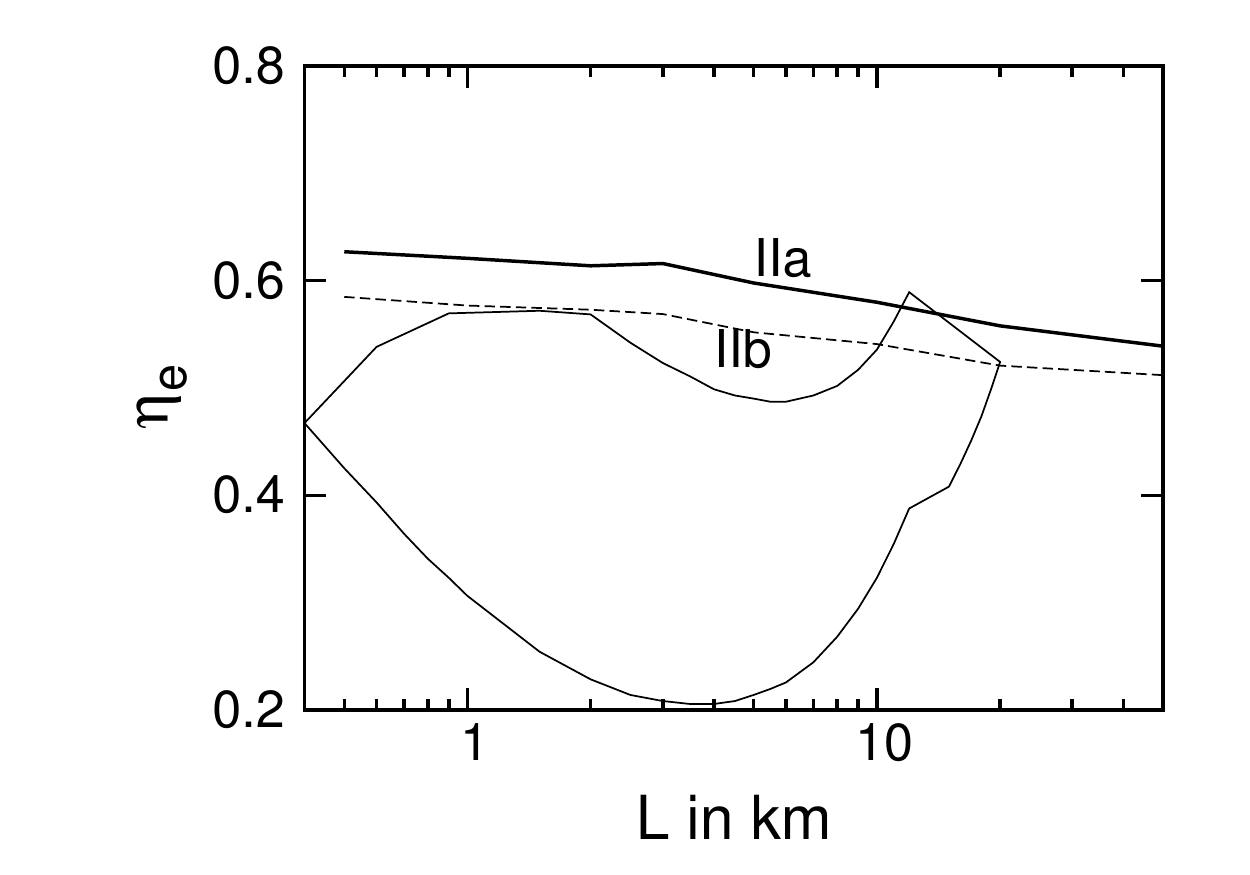} \\
\end{tabular}$%\vspace{-0.3cm}
\caption{$\eta_e$ as a function of the coronal width $L$ for the
accretion disk-corona geometry. Here the extent of the disk
$R_{max}-R_{min}$ = 10 kms. The left  and right panel are for the hot 
and cold seed photon models.  The solid and dashed lines
are for two corresponding spectral parameters. The closed curves
show the allowed range of $\eta$ and size obtained by \citep{Kumar-Misra2016}      
}
\label{dketa}
\end{figure}

\subsection{Disk-Corona  geometry} \label{DK:}

We next consider a third possible geometry, 
i.e., of an optically thick accretion disk sandwiched by an hot corona as
shown in the right panel of Figure \ref{blgeo}. The height of the corona
is taken to be $L$ while the disk and the corona above
 is considered to span from an inner radius of of $R_{min}$ to $R_{max}$.
For computational purposes we have introduced a thickness of the disk
of $R_s = 0.2$ kms but the results, as expected, are insensitive to this value.
In fact the determining parameter here is the ratio of the height of
the corona to the annular width of the disk $R_{max} - R_{min}$. Thus
we   fix $R_{max} - R_{min} = 10$ kms and vary $L$. 

In Figure \ref{dketa}, we plot the fraction $\eta_e$ versus height $L$ for
different spectral parameter values and as before compare with ranges
obtained in Paper I. As expected, there is only a weak dependence of $\eta_e$
on $L$ and it has  a rather large value of $\sim 0.7$.
In fact for the ``cold'' seed photon case $\eta_e$ is marginally larger
than the maximum value obtained in Paper I. This seems to suggest
that for this case at least, such a disk-corona geometry is unfavourable.
However, given the large uncertainties it is difficult to make concrete
statements. Nevertheless, our results show that for such a geometry
the value of $\eta_e$ is expected to be large, more or less independent of
the thickness of the corona.

\section{Summary and Discussion}

Using a Monte Carlo scheme, we estimate the fraction of Comptonized
photons that impinge back into the seed photon source $\eta$ for
different geometries and spectral parameters relevant to neutron star
low mass X-ray binaries. The primary motivation was that to explain
the observed soft lags in KHz QPOs, one needed to invoke a large value
$\eta$ in the range of $0.2$-$0.6$  and it was important to find out if  
this range can be achieved for any reasonable accretion geometry.

We consider three kinds of geometries for the Comptonizing medium 
 which are (i) a spherical shell around the neutron
star, (ii) a boundary layer system where the medium is taken to be
a rectangular torus around the star and (iii) a corona sandwiching
a thin accretion disk. We consider different sizes for the medium and
a range of spectra parameters. In particular we consider two extreme
cases of spectral parameters for the 
two degenerate spectral models which are called
the hot and cold seed photon models. 

Our basic result is that for a wide range of reasonable sizes and
spectral parameters, the values of $\eta_e$ computed by the Monte Carlo method
lie within $0.2$ to $0.8$ and hence are compatible with the values used
by \citet{Kumar-Misra2016} to explain the soft time lags of the kHz QPOs.
Since the range of $\eta$ and size inferred from fitting the 
time-lags are rather broad, we cannot concretely rule out any of the
 three geometries considered. However, it seems that the boundary layer
geometry can have $\eta$ values more in line with what is required and
the disk-corona geometry produces $\eta$ values which are marginally larger.
Our results show that it is possible to constrain the geometry of the
system if high quality data for energy dependent time-lags are available.
We look forward to data from the recently launched satellite
{\it ASTROSAT}\footnote{http://astrosat.iucaa.in}\citep{Agrawal2006, Singh-etal2014}, which might provide such high quality data. Perhaps it would then
be warranted to consider other complexities such as the seed photons for
the boundary layer case may be produced in the accretion disk rather than
the neutron star surface or that the corona on top of the accretion disk maybe
in the form of inhomogeneous clumps rather than being a uniform medium. Also,
 not all the photons that impinge back into the source, will be absorbed 
and one needs to solve the  radiative transfer equations self consistently 
to find the fraction reflected. This reflected emission will have light travel
time delays which may significantly effect the time-lags between different
energy bands.

Finally, it is interesting to note that for the geometries considered here a
significant fraction of the photons impinge back into soft photon source.
This effect needs to be taken into account in any detailed study of 
these X-ray binaries.

%\clearpage  
\section*{ACKNOWLEDGEMENTS}
NK thanks CSIR/UGC for providing support for this work.

\label{lastpage}

\end{document}